\documentstyle[preprint,aps]{revtex}

\begin{document}
\title{Generalized Nonlinear Equation and Solutions for Fluid Contour
Deformations}
\author{A. Ludu}
\address{Dept. Chemistry and Physics, Northwestern State University,\\
Natchitoches, LA 71497}
\author{A. R. Ionescu}
\address{Institute for Atomic Physics, Bucharest, Romania}
\date{\today }
\maketitle
\pacs{05.45.-a, 43.35.+d, 83.10.Ji, 47.55.Dz, 43.25.Rq, 47.35.+i}

\begin{abstract}
We generalize the nonlinear one-dimensional equation for a fluid layer
surface to any geometry and we introduce a new infinite order differential
equation for its traveling solitary waves solutions. This equation can be
written as a finite-difference expression, with a general solution that is a
power series expansion with coefficients satisfying a nonlinear recursion
relation. In the limit of long and shallow water, we recover the Korteweg-de
Vries equation together with its single-soliton solution.
\end{abstract}

\section{Introduction}

Shape deformations are important for an understanding of diverse many-body
systems like the dynamics of suspended liquid droplets [1], long-lying
excitations of atomic nuclei or ''rotation-vibration'' excitations of
deformed nuclei and their fission or cluster emission modes [2]
hydrodynamics of vortex patches [3], evolution of atmospheric plasma clouds
[4], formation of patterns in magnetic fluids and superconductors [5],
electronic droplets and quantum Hall effect [6], as well as resonant
formation of symmetric vortex waves [7], etc. \ 

The theoretical description of such systems is often realized in terms of \
collective modes, such as large amplitude collective oscillations in nuclei,
sound waves in solids, collective excitations in BEC, plasmons in charged
systems, or surface moves in {\it tubifex} worms populations. Collective
modes are especially important when their energies are lower than competing
singe-particle degrees of freedom. Sometimes, however, single-particle or
collective modes in the bulk of a system show particularly dense or sparse
spectra. Systems of latter type are often referred as incompressible
[1,5-7]. The incompressibility can be hard or relatively soft and still
serves as a convenient limit for describing large differences in relevant
length or time scales, as for the macroscopic motion of a liquid drop. Under
these conditions, one can usually focus attention on the motion of the
boundaries of the system, which usually has softer modes with frequencies
that are lower than those of the bulk (e.g. Rayleigh waves for a solid or
surface waves for a liquid drop travel at speeds considerable slower than
the bulk sound waves).

Concentrating on the motion of the boundary of the system has considerable
advantage: a reduction in the dimensionality of the problem that leads to
relatively simple analytical treatments or a tremendous reduction in the
effort needed to numerically solve or simulate the problem. Furthermore,
associated with the incompressibility one usually finds microscopic
conservation laws that expand into global constrains on the whole systems,
even when the microscopic dynamics is completely local, as for example when
the volume of a liquid drop is conserved.

These shape deformations and their dynamics have played an important role in
gaining a better understanding of numerous problems in diverse fields of
physics. The incompressibility is normally reflected in the existence of a
field that is piecewise constant, so that there is a sharp boundary between
two or more distinct regions of the space with different physical
properties. This field can be of classical origin like the density of a
liquid or the charge density of a plasma, or it can originate in the
quantum-mechanical properties of the system, like the magnetization of a
type-II superconductor. Among the various examples where these questions are
relevant, the waves on the surface of a liquid represent a unique
opportunity to study the dynamics of shape deformations in a {\it clean} and 
{\it controlled} environment. We showed recently [1] that the surface
dynamics of an incompressible liquid drop can be modeled by the Korteweg-de
Vries (KdV) or the modified KdV (mKdV) equations for the circular shape
[8,9]. The KdV equation was proposed a century ago, [9], to explain the
dynamics of water waves observed in a channel by Scott Russel in 1834.

The main property of this equation is the near-equal importance of nonlinear
and dispersion effects. Some of its most important solutions, like solitary
waves, are single, localized, and non-dispersive structures that have a
localized finite energy density. Among these solutions, the solitons are the
solitary waves with special added requirements concerning their behavior at
infinity ($x\rightarrow \infty $, $t\rightarrow \infty $) and having special
properties associated with scattering with other such single-solitary wave
solutions. Despite this nonlinearity, the KdV equation is an infinite
dimensional Hamiltonian system [10] and, remarkably, when the KdV solution
evolves in time, the eigenvalues of the associated Sturm-Liouville operator $%
{\partial }^{2}+\eta $ (where $\eta $ is the solution of the KdV equation)
remain constant. Several extended and complex methods have been developed to
study and to solve the KdV equation and other nonlinear wave equations like
the nonlinear Schr\"{o}dinger equation (NLS), the sine-Gordon equation (SG),
etc. Among these there is the inverse scattering theory (IST), theoretical
group methods, numerical approaches. For a review of some of these
techniques one can look into the book by Witham [11] or the paper by Scott 
{\it et al} [12].

Due to its properties the KdV equation, and its quadratic extension mKdV,
were the source of many applications and results in a large area of
non-linear physics (for a recent review see [13] and the references herein).
However, in all these applications, the KdV equation and its various
generalizations (mKdV, KdV hierarchy, KP equation, supersymmetric
generalizations, etc.) appear as the consequence of certain simplifications
of the physical systems, especially due to some perturbation techniques
involved.

In the present paper we use only one of the two well-known necessary
conditions for obtaining the KdV equation from a one-dimensional shallow
water channel, i.e.only the smallness of the amplitude of the soliton $\eta
_{0}$ with respect to the depth of the channel, $h$. This is the single
condition which we use, with $h$ taken to be an arbitrary parameter. In
order to obtain the KdV equation from shallow water channel models, a second
condition is imposed, i.e. the depth $h$ of the channel should be smaller
than the half-width $\lambda $ of the solitary wave. In the following, in
our generalized approach, we do not impose this second condition nor the
assumption of an infinite length channel; the channel length $L$, is taken
to be an arbitrary parameter. By doing this we are actually investigating
exact solutions for nonlinear modes in finite domains. What we actually do
is replace the fixed-length boundary condition by a periodic requirement and
in this way model general surface dynamics equations for liquid drops or
liquid drop-like systems. This generalizations is important in connection
with atomic and nuclear physics applications where the height of the
perturbation may be comparable to the atomic or nuclear radius (big clusters
formation on the surface of symmetric molecules or nuclear molecules) so the
dynamics of a shallow fluid layer becomes inappropriate. Therefore, we
study, as a first step, the non-linear dynamics of a fluid of arbitrary
depth in a bounded domain. This different starting point leads us to a new
type of equation which generalizes in some sense the KdV equation (higher
order in the derivatives and higher order nonlinearity) and also reduces to
it the shallow liquid case.

As we shall show, this generalized result (infinite order differential
equation and higher order nonlinearity) can also be written in terms of a
finite-difference equation. In this context we introduce here another
possible physical interpretation for the translation operator in fluid
dynamics, by relating its associate parameter to the depth of the fluid
layer. In the present model we take into account not only the existence of a
uniform force field (like an electric field or the gravitational field) but
also the influence of the surface pressure acting on the free surface of the
fluid. This implies that the KdV-like structure for the dynamically equation
arrises in first order in the smallness parameter and one does not have to
rely on a second order effect to see its consequences on the dynamics. We
also note that, despite the general tendency followed by papers concerning
applications of differential-finite-difference equations in physics of first
introducing the equation and then searching possible applications, we
obtained an infinite order partial differential equation and its connections
to a finite-difference equation, in a natural way, starting from a
traditionally physical one-dimensional hydrodynamic model.

\section{The generalized KdV equation}

Consider a one-dimensional ideal incompressible fluid layer with depth $h$
and constant density $\rho $, in an uniform force field. We suppose
irrotational motion and consequently the velocity field is obtained from a
potential function $\Phi (x,y,t),$ that is $\vec{V}(x,y,t)=\nabla \Phi
(x,y,t)$ where we denote the two components of the velocity field $\vec{V}%
=(u,v)$. The continuity equation for the fluid results in the Laplace
equation for $\Phi $, $\Delta \Phi (x,y,t)=0$. This Laplace equation should
be solved with appropriate boundary conditions for the physical problem of
interest. We take a 2-dimensional domain: $x\in \lbrack x_{0}-L,x_{0}+L]$
(as the ''horizontal'' coordinate) and $y\in \lbrack 0,\xi (x,t)]$ (as the
''vertical'' coordinate), where $x_{0}$ is an arbitrary parameter, $L$ is an
arbitrary length, and $\xi (x,t)$ is the shape of the free surface of the
fluid. The boundary conditions on the lateral walls $x=x_{0}\pm L$ and on
the bottom $y=0$ consist of the vanishing of the normal component of the
velocity. The free surface fulfills the kinematic condition

\begin{equation}
v|_{\Sigma }=({\xi }_{t}+{\xi }_{x}u)|_{\Sigma },
\end{equation}%
where we denote by $\Sigma $ the free surface of equation $y=\xi (x,t)$ and
the subscript indicates the derivative. Eq. (1) expresses the fact that the
fluid particles which belong to the surface remain on the surface during
time evolution. By taking into account the above boundary conditions on the
lateral walls and on the bottom, we can write the potential of the
velocities in the form:

\begin{equation}
\Phi (x,y,t)=\sum_{k=0}^{\infty }\cosh \left( {\frac{{k\pi y}}{L}}\right)
\left( {\alpha }_{k}(t)\cos \left( \frac{{k\pi x}}{L}\right) +{\beta }%
_{k}(t)\sin \left( \frac{{k\pi x}}{L}\right) \right) ,
\end{equation}%
where ${\alpha }_{k}$ and ${\beta }_{k}$ are time dependent coefficients
fulfilling the condition

\[
{\alpha }_{k}\sin {\frac{{k\pi (x_{0}\pm L)}}{{L}}}={\beta }_{k}\cos {\frac{{%
k\pi (x_{0}\pm L)}}{{L}}}, 
\]%
for any positive integer $k$. This restriction introduces a special time
dependence of $\ \alpha _{k}$ and ${\beta }_{k}$, i.e. $\frac{\alpha _{k}}{%
\beta _{k}}=\gamma _{k}=$constant for any $k$, $\beta \neq 0$. If ${\beta }%
_{k}=0$ we simply equate with $0$ the inverse of the above fraction. The $%
\alpha _{k}(t)$ and ${\beta }_{k}(t)$ functions also depend on $x_{0}$. This
special time dependence doesn't affect the general nature of the potential $%
\Phi $, but does affect the balance between the two terms in the RHS of eq.
(2). If we set all ${\beta }_{k}=0,$ we can take arbitrary values for $x_{0}$
and general form for ${\alpha }_{k}$. In the infinite channel limit, $%
L\rightarrow \infty $, there are no more restrictions concerning ${\alpha
_{k}}$ and ${\beta }_{k}$ functions, and we can choose $x_{0}=0$ without
loss of generality. We introduce the function:

\begin{equation}
f(x,t)=\sum_{k=0}^{\infty }\left( -{\alpha }_{k}(t)\sin \left( k\pi {\frac{x%
}{L}}\right) +{\beta }_{k}(t)\cos \left( k\pi {\frac{x}{L}}\right) \right) {%
\frac{{k\pi }}{L}},\hfill
\end{equation}%
so the velocity field can be written like:

\begin{eqnarray}
u &=&{\Phi }_{x}=\cos (y\partial )f(x,t)  \nonumber \\
v &=&{\Phi }_{y}=-\sin (y\partial )f(x,t).\hfill
\end{eqnarray}%
where, for simplicity, the operator $\partial $ represents the partial
derivative with respect to the $x$ coordinate. Equations (4) do not depend
on $L$ and therefore any approach toward the long channel limit must include
the $L\rightarrow \infty $ (unbounded) limit. In the following we describe
excitations of small height \ compared to the depth, and not necessarily
large widths. Also, in the boundary condition eq.(1)we use velocities
evaluated at $y=\xi (x,t)=h+\eta (x,t)$ to the first order in $\eta $.
Eqs.(4) reads:

\begin{eqnarray}
u(x,\xi (x,t),t) &=&\left[ \cos (h\partial )-\eta (x,t)\partial \sin
(h\partial )\right] f(x,t)  \nonumber \\
v(x,\xi (x,t),t) &=&-\left[ \sin (h\partial )+\eta (x,t)\partial \cos
(h\partial )\right] f(x,t).\hfill
\end{eqnarray}%
The dynamics of the fluid is described by the Euler equation at the free
surface. The equation that results is written on the surface $\Sigma $ it in
terms of the potential and differentiated with respect to $x$. By imposing
the condition $y=\xi (x,t),$ and by using a constant force field we obtain
the form%
\begin{equation}
u_{t}+uu_{x}+vv_{x}+g{\eta }_{x}+{\frac{1}{{\rho }}}P_{x}=0,\hfill
\end{equation}%
where $g$ represents the the force field constant and $P$ is the surface
pressure. Following the same approach as used in the calculation of surface
capillary waves [17], we have for our one-dimensional case

\begin{equation}
P|_{\Sigma }={\frac{{\sigma }}{{\cal R}}}={\frac{{\sigma {\eta }_{xx}}}{{{(1+%
{\eta }_{x}^{2})}^{3/2}}}}\simeq -\sigma {\eta }_{xx},\quad \quad \text{for
small }\eta ,
\end{equation}%
where ${\cal R}$ is the local radius of curvature of the surface (in this
case, the curvature radius of the curve $y=\xi (x,t)$) and $\sigma $ is the
surface pressure coefficient. Inside the fluid the pressure is given by the
Euler equation. The nonlinearities appear in the dynamics through the
nonlinear terms in eqs. (1), (5), (6) and (7). Consequently, we have a
system of two differential equations (1 and 6) for the two unknown
functions: $f(x,t)$ and $\eta (x,t)$, with $u$ and $v$ depending on $\eta $
and $f$ from eqs.(5). With $f$ and $\eta $ determined and introduced in the
expression for $u$ $\ $and $v,$we can finally find the coefficients ${\alpha 
}_{k}$ and $\beta _{k}$. In the following we treat these equations in the
approximation of small perturbations of the surface $\Sigma $, with respect
to the depth, $a=max|{\eta }^{(k)}(x,t)|<<h$, where $k=0,...,3$ are the
orders of differentiation. In the linear approximation eqs. (1) and (6)
become, respectively

\begin{eqnarray}
-\sin (h\partial )f &=&{\eta }_{t}  \nonumber \\
\cos (h\partial )f_{t} &=&-g{\eta }_{x}+{\frac{{\sigma }}{{\rho }}}{\eta }%
_{xxx},\hfill
\end{eqnarray}%
and we obtain, by eliminating ${\xi }$ from eqs.(8)

\begin{equation}
\cos (h\partial ){\eta }_{tt}=\sin (h\partial )\left( {\frac{{c_{0}^{2}}}{{%
h^{2}}}}{\eta }_{x}-{\frac{{\sigma }}{{\rho }}}{\eta }_{xxx}\right) ,\hfill
\end{equation}%
which, in the lowest order of approximation in $(h\partial )$ for the sine
and cosine functions, and in the absence of the surface pressure, gives us
the familiar wave equation ${\eta }_{tt}=c_{0}^{2}{\eta }_{xx}$, where $%
c_{0}=\sqrt{gh}$ is the sound wave velocity. By introducing the solution $%
\eta =e^{i(kx-\omega t)}$ in the linearized eq.(9) we obtain a nonlinear
dispersion relation

\begin{equation}
\omega ^{2}=c_{0}^{2}k^{2}\left( 1+\frac{\sigma }{\rho g}\right) \frac{\tanh
\left( kh\right) }{kh}.
\end{equation}%
In the limit of shallow waters we find for the dispersion relation, in the $%
\sigma \neq 0$ case of an acoustical wave as well as the $\sigma >>\rho g$
surface capillary wave limit, $\omega ^{2}=\frac{h\sigma }{\rho }k^{4}$. In
the absence of the surface pressure ($\sigma =0$) the function $f$ is given
in this linear approximation, at least formally, by

\begin{equation}
f^{lin}=\frac{c_{0}}{h}\left( \frac{\sin \left( 2h\partial \right) }{%
2h\partial }\right) ^{-\frac{1}{2}}\eta ,
\end{equation}%
which in the limit of a shallow fluid has a particular solution of the form

\begin{equation}
f^{0}(x,t)=\frac{c_{0}}{h}\eta .
\end{equation}%
For the time derivative of $f$ we have, from the second equation of eq.(8)

\begin{equation}
f_{t}^{lin}(x,t)=\left( \cos \left( h\partial \right) \right) ^{-1}\left( -%
\frac{c_{0}^{2}}{h}\eta _{x}+\frac{\sigma }{\rho }\eta _{xxx}\right) ,
\end{equation}%
which in the limit of a shallow fluid reduces to

\[
f_{t}^{0}(x,t)=-\frac{c_{0}^{2}}{h}\eta _{x}+\frac{\sigma }{\rho }\eta
_{xxx}. 
\]%
Following [10] we look for the solution of eqs.(1 and 6) in the form

\[
f=\frac{a}{h}c_{0}\tilde{\eta}+\left( \frac{a}{h}\right) ^{2}f_{2}, 
\]

\[
f_{t}=-c_{0}^{2}\frac{a}{h}\left( \cos \left( h\partial \right) \right) ^{-1}%
\tilde{\eta}_{x}+\frac{a\sigma }{\rho }\left( \cos \left( h\partial \right)
\right) ^{-1}\tilde{\eta}_{xxx}+\left( \frac{a}{h}\right) ^{2}g_{2}, 
\]%
which represents a sort of perturbation technique in $\frac{a}{h}$, where $%
\eta =a\tilde{\eta}$. Of course a functional connection exists between the
perturbation $f_{2}(x,t)$ and $g_{2}(x,t)$. Eq.(1) yelds in the lowest order
in $\frac{a}{h}$

\begin{equation}
-c_{0}\sin \left( h\partial \right) \tilde{\eta}=h\tilde{\eta}_{t}+ac_{0}(%
\tilde{\eta}_{x}\cos \left( h\partial \right) \tilde{\eta}+\tilde{\eta}\cos
\left( h\partial \right) \tilde{\eta}_{x}).
\end{equation}%
If we approximate $\sin (h\partial )\approx h\partial -\frac{1}{6}\left(
h\partial \right) ^{3}$, $\cos (h\partial )\approx 1-\frac{1}{2}\left(
h\partial \right) ^{2}$, we obtain from eq.(14) the polynomial differential
equation

\begin{equation}
a\tilde{\eta}_{t}+2c_{0}\epsilon ^{2}h\tilde{\eta}\tilde{\eta}%
_{x}+c_{0}\epsilon h\tilde{\eta}_{x}-c_{0}\epsilon \frac{h^{3}}{6}\tilde{\eta%
}_{xxx}-\frac{c_{o}\epsilon ^{2}h^{3}}{2}\left( \tilde{\eta}_{x}\tilde{\eta}%
_{xx+}\tilde{\eta}\tilde{\eta}_{xxx}\right) =0,
\end{equation}%
where $\epsilon =\frac{a}{h}$. The first four terms in eq.(15) correspond to
the zeroth approximation for terms in $f,$obtained from boundary conditions
at the free surface, in eq.(6.1.15a) from Lamb's book [10], i.e., the
traditional way of obtaining the KdV equation in shallow channels. In this
case all the terms are first and second order in $\epsilon $. If we apply
the second restriction with respect to the solution, that is the half-width
is much larger than $h$, we can neglect the last parenthesis in eq.(15) and
we obtain exactly the KdV equation for the free surface boundary conditions.
In other words we understand the condition $\ h\partial $ is ''small'' like $%
\left( h\partial \right) f\left( x,t\right) <<1$ over the entire domain of
definition of$f$. This means that the spatial extension of the perturbation $%
f(x,t)$ is large compared to $h$, which is exactly the case in which the KdV
equation arrises from the shallow water model (see Chapter 6 in [3], $%
h\partial f(x,t)$ of order $\approx \frac{h}{L}=\delta <<1$).

By using again the approximations given by eq.(5) we can write the Euler
equation eq.(6) in the form

\begin{equation}
\partial _{t}\Omega f+\Omega (\partial \Omega f)+\Omega f(\partial
_{t}\Omega f)+\omega f(\partial _{t}\Omega f)=-g\eta _{x}+\frac{\sigma }{%
\rho }\eta _{xxx},
\end{equation}%
where we use the notation

\begin{eqnarray}
\Omega &=&\cos (h\partial )-\eta \partial \sin (h\partial ), \\
\omega &=&\sin (h\partial )+\eta \partial \cos (h\partial ).  \nonumber
\end{eqnarray}%
Note that the operators given in eq.(17) satisfy the following interesting
relation

\begin{equation}
\Omega +i\omega =e^{i\partial }|_{\Sigma }+{\cal O}_{2}(h\partial )+{\cal O}%
_{2}(h\eta ).
\end{equation}%
Following the same procedure as for the free surface boundary condition
eq.(3) namely using an approximation for small $\eta $ we obtain from eq.(16)

\begin{equation}
\cos (h\partial )f_{t}=-g\eta _{x}+\frac{\sigma }{\rho }\eta _{xxx},
\end{equation}%
which, in the lowest order in $\frac{a}{h}$ and by using eq.(12 and 13)
reduces to an identity.

Before further analysis we would like to note that in the shallow water
case, following the same notation as in Chapter 6 of Lamb [10], that is $\
\epsilon =\frac{a}{h},$ $\delta =\frac{h}{l}$, where $l$ gives the order of
magnitude of the half-width of the perturbation \ $\eta $, and introducing a
new parameter $\ \alpha =-\frac{\sigma }{gl^{2}\rho }$, we obtain for the
Euler equation the form

\begin{equation}
\tilde{\eta}_{t^{\prime }}+\tilde{\eta}_{x^{\prime }}+\frac{3}{2}\tilde{\eta}%
\tilde{\eta}_{x^{\prime }}+\alpha \frac{\epsilon }{2}\tilde{\eta}_{x^{\prime
}x^{\prime }x^{\prime }}=0,
\end{equation}%
which is again the KdV equation. The primes attached to the subscripts
denote dimensionless units [10]. The difference between eq.(20) and the
corresponding eq.(1.15.b) from Lamb's book is given by the inclusion of the
surface pressure effects. If the coefficient $\alpha $ is large enough one
can ignore second order terms, like for example $\delta ^{2}$, to obtain the
KdV equation. Of course, the above approach introduces changes in the
differential equations which involve higher order perturbations like $%
f^{(1)} $ and $f^{(2)}$ from [10] or $f_{2}$ , $g_{2}$ in our case. \ The
reduction of eq.(1) to KdV equation occurs if in eq.(14) we limit the
expression to terms that are at most third order in $\ h\partial $

\begin{equation}
\eta _{t}+c_{0}\eta _{x}-c_{0}\frac{h^{2}}{6}\eta _{xxx}+\frac{2c_{0}}{h}%
\eta \eta _{x}=0.
\end{equation}

In the following we shall investigate this generalized KdV equation (gKdV)
obtained from eq.(14) by keeping all terms in sin and cos, namely

\begin{equation}
\eta _{t}+\frac{c_{0}}{h}\sin (h\partial )\eta +\frac{c_{0}}{h}(\eta
_{x}\cos (h\partial )\eta +\eta \cos (h\partial )\eta _{x})=0.
\end{equation}%
Eqs.(1) and (16) yield, to higher orders in $\frac{a}{h}$ and in $h\partial $%
, corresponding differential equations for the functions $f_{2}$ and $g_{2}$%
, but here we shall study only eq.(21).

\section{The finite-difference form of the gKdV equation}

In this section we limit ourselves to the steady-state translational waves
and consider only solutions of the form $\eta (x,t)=\eta (x+Ac_{0}t)=\eta
(X) $ where $A\in {\cal R}$ and $X=x+Ac_{0}t$. Eq.(21) can be written in the
form

\[
Ah\eta _{X}(X)+\frac{\eta (X+ih)-\eta (X-ih)}{2i}+\eta _{X}(X)\frac{\eta
(X+ih)+\eta (X-ih)}{2} 
\]

\begin{equation}
+\eta (X)\frac{\eta (X+ih)+\eta (X-ih)}{2}=0,
\end{equation}%
if we suppose that $\eta $ is an analytic function in the domain $%
\mathop{\rm Re}%
(z)\in (-\infty ,\infty ),$ $%
\mathop{\rm Im}%
(z)\in (-h,h),$ $z=x\pm ih$ of the complex plane. We study rapidly
decreasing solutions at infinity and we make the substitution $v=e^{Bx}$ for 
$x\in (-\infty ,0)$ and $v=e^{-Bx}$ for $x\in (0,\infty ),$where $B$ is a
positive constant. By introducing $\eta (X)=-hA+f(v)$ we obtain a
differential-finite-difference equation (DFDE) for the function $f(v)$

\begin{equation}
f(v)\frac{\delta f_{v}^{2}(v)}{\delta f_{v}(v)}+f(v)\frac{\delta f^{2}(v)}{%
\delta f(v)}+2\frac{\sin (Bh)}{B}\delta f(v)=0,
\end{equation}%
where we define the finite-difference operator as

\begin{equation}
\delta f(v)=\frac{f(e^{iBh}v)-f(e^{-iBh}v)}{e^{iBh}v-e^{-iBh}v}.
\end{equation}%
We can write the solution of eq.(23) (or eq.(24)) as a power series in v

\begin{equation}
f(v)=\sum_{n=0}^{\infty }a_{n}v^{n},
\end{equation}%
and we choose $a_{0}=hA$ in order to have $\lim_{x\rightarrow \pm \infty
}\eta (x)=0.$ Equation (25) results in a non-linear recursion relation for
the coefficients $a_{n}$, that is

\[
\left[ Ahk+\frac{1}{B}\sin \left( Bhk\right) \right] a_{k} 
\]

\begin{equation}
=-\sum_{n=1}^{k-1}n\left( \cos \left( Bh\left( k-n\right) \right) +\cos
\left( Bh\left( k-1\right) \right) \right) a_{n}a_{k-n}.
\end{equation}%
By taking $k=1$ in the above relation, we obtain $a_{1}\left[ Ah+\frac{1}{B}%
\sin \left( Bh\right) \right] =0$. Without loss of generality

and because of the arbitraryness of $B$ we can write

\begin{equation}
A=-\frac{\sin \left( Bh\right) }{Bh}.
\end{equation}%
This relation fixes the velocity of the envelope of the perturbation if its
asymptotic behavior is fulfilled. In order to have $A\neq 0$, we need $%
Bh\neq k\pi $ for $k$ integer. In this \ condition $\ a_{1}$ is still
arbitrary and by writing $a_{k}=\alpha _{k}a_{1}^{k}$ we have $\alpha _{1}=1$
and the recursion relation

\begin{equation}
\alpha _{k}=\frac{2B\cos \frac{Bh(k-1)}{2}}{k\sin \left( Bh\right) -\sin
\left( kBh\right) }\sum_{n=1}^{k-1}n\cos \frac{Bh\left( 2k-n-1\right) }{2}%
\alpha _{n}\alpha _{k-n},
\end{equation}%
for $k\geq 2$. This recursion\ relation gives the coefficient for $k$ in
terms of those for $k-1$ and lesser values. For a smooth behavior of the
solution $\eta (X)$ at $X=0$, i.e. continuity of its derivative, we must
introduce the condition

\begin{equation}
f_{v}(1)=\sum_{n=1}^{\infty }n\alpha _{n}a_{1}^{n-1}=0,
\end{equation}%
or require that the derivative of the power series $f(v)$ with coefficients
given in eq.(28) to be zero in $z\in R,$ $z=a_{1}.$ This set the value for $%
a_{1}.$

In the following we study a limiting case of the relation eq.(28), by
replacing the $sin$ and $cos$ \ expressions with their lowest nonvanishing
terms in their power expansions

\begin{equation}
\alpha _{k}=\frac{6}{B^{2}h^{3}k\left( k^{2}-1\right) }\sum_{n=1}^{k-1}n%
\alpha _{n}\alpha _{k-n.}
\end{equation}%
It is straightforward exercise to prove that

\begin{equation}
\alpha _{k}=\left( \frac{1}{2B^{2}h^{3}}\right) ^{k-1}k,
\end{equation}%
is a solution of the recursion equation. This can be done using mathematical
induction and by taking into account the relations

\begin{eqnarray*}
\sum_{n=1}^{k-1}n^{2} &=&\frac{k(k-1)(2k-1)}{6} \\
\sum_{n=1}^{k-1}n^{3} &=&\left( \frac{k(k-1)}{2}\right) ^{2}.
\end{eqnarray*}%
We can write the power expansion

\begin{equation}
g(z)=\sum_{k=1}^{\infty }k\left( \frac{1}{2B^{2}h^{3}}\right) ^{k-1}z^{k},
\end{equation}%
which has the radius of convergence ${\cal R}=2B^{2}h^{3}$ (due to the
Cauchy-Hadamard criteria). The function $g(z)$ can be written in the form

\begin{equation}
g(z)=z\left( \frac{1}{1-\frac{z}{2B^{2}h^{3}}}\right) _{z}2B^{2}h^{3}=-\frac{%
z}{\left( 1-\frac{z}{2B^{2}h^{3}}\right) ^{2}}.
\end{equation}%
Condition eq.(31) results in $a_{1}=-2B^{2}h^{3}$ and

\begin{equation}
\alpha _{k}=k\left( \frac{1}{2B^{2}h^{3}}\right) ^{k-1}\left(
-2B^{2}h^{3}\right) ^{k}=2B^{2}h^{3}\left( -1\right) ^{k},
\end{equation}%
which provides

\begin{eqnarray}
\eta (x) &=&2B^{2}h^{3}\sum_{k=1}^{\infty }k\left( -e^{-B\left| X\right|
}\right) ^{k}  \nonumber \\
2B^{2}h^{3}\frac{e^{-B\left| X\right| }}{\left( 1+e^{-B\left| X\right|
}\right) ^{2}} &=&  \nonumber \\
&&\frac{B^{2}h^{3}}{2}\frac{1}{\left( \cosh \left( \frac{BX}{2}\right)
\right) ^{2}}.
\end{eqnarray}%
As expected, this solution is exactly the single-soliton solution of the KdV
equation and it was indeed obtained by assuming $h$ small in the recursion
relation eq.(28). So, the gKdV equation has two general features: in the
reduction given by eq.(16) it yields the KdV equation, and for eq.(230,
namely in the limit $h\partial $ ''small'' the differential equation and one
of its solutions, $\eta (X)$ go into the KdV equation and its single-soliton
solution.

In general we do not obtain a simple solution like the relation (33) but we
have all the necessary information from eq.(28). It seems that the power
series $g(z)$ with the coefficients given by the recursion relation (28) has
a nonvanishing radius of convergence. The problem of the existence of a real
point $z_{o}$ in the disk of convergence or on its edge ($g^{\prime
}(z_{o})=0$) needs further study. For the KdV equation, this point is on the
edge of the disk of convergence of the power series with coefficients given
by the recursion relation (28). We mention that $g^{\prime }(z_{o})=0$ means
that the function is not univalent (it is not injective in the neighborhood
of this point).

An interesting observation concerning a conjecture formulated by Bieberbach
in 1916 seems to be worth noting: if $f(z)=z+a_{2}z^{2}+\dots $ is analytic
and univalent in the unit disk, then $\left| a_{n}\right| \leq n$ for all $n$%
, with equality occuring only for rotations of the Koebe function $k(z)=%
\frac{z}{\left( 1-z\right) ^{2}}$ [14]. The form of our solution for the KdV
equation, eq.(34), is exactly the Koebe function up to a scaling of the
variable $z$. This conjecture was proved in 1986 by de Branges [15]. This
fact suggests that one could obtain information about the possible zeros of
the $g^{\prime }(z)$ if it is possible to obtain estimates for the radius of
convergence of the power series with coefficients given by eq.(30).

The change of the variable that we used to obtain the nonlinear DFDE eq.(25)
can be used in the time-dependent equation obtained from eq.(23). If $\eta
(x,t)=f(v,t)$ where $v=e^{Bx}$ or $v=e^{-Bx}$, then we obtain

\begin{eqnarray}
&&2\frac{h}{c_{0}}\frac{1}{v}f_{t}^{\mp }(v,t)\pm Bf_{v}^{\mp }(v)\frac{%
\delta f^{\mp 2}(v,t)}{\delta f^{\mp }(v,t)}  \nonumber \\
\pm Bf^{\mp }(v)\frac{\delta f^{\mp 2}(v,t)}{\delta f^{\mp }(v,t)}\pm 2\sin
\left( Bh\right) \delta f^{\mp }(v,t) &=&0,
\end{eqnarray}%
where $\pm $ refers to the positive or negative real semiaxis. If we try a
solution of the form $f^{\pm }(v,t)=\sum_{k=0}^{\infty }a_{k}^{\pm }v^{k}$,
we obtain the following infinite system of ODE:

\begin{eqnarray}
\frac{f}{c_{0}}\frac{d}{dt}a_{k}^{\mp }(t) &=&\pm \sin \left( kBh\right)
a_{k}^{\mp }(t)  \nonumber \\
&&\mp 2B\cos \frac{Bh(n-1)}{2}\sum_{n=0}^{k}na_{n}^{\mp }(t)a_{k-n}^{\mp
}(t)\cos \left( \frac{\left( 2k-n-1\right) Bh}{2}\right) .
\end{eqnarray}%
We note that if $a_{0}(t_{0})=0$ this relation is preserved in time. The
above system can be resolved step-by-step when the equation for $a_{n}$
involves only the coefficients $a_{k}$ with $k\leq n$. However, the smooth
behavior of the solution in $x=0$ \ at any moment of time is a nontrivial
problem.

\section{Conclusions}

We have shown that the KdV equation used to describe fluids in shallow
channels can be generalize to liquids flowing at any depth or length. As a
consequence the role of the depth parameter in various applications that use
such a model can be explored, like for example the dynamics of cluster
formation and fission. The present paper announces two key results: a
generalization of the KdV equation starting from a physical model; and
anembeding of this nonlinear equation into a DFDE.

By using a nonlinear hydrodynamic approach for a fluid layer of arbitrary
depth and length, we obtained a differential equation of infinite order
which is the generalization of the KdV equation for the corresponding
shallow water pattern. We succeeded in rewriting this equationin the form of
an finite-difference multiscale formalism, and consequently we have obtained
a nonlinear recursion relation for the coefficients of its general traveling
solution. We stress the importance of the introduction of the surface
pressure term, eqs.(8,9) which yields the dispersion term, that is the term
proportional to $\eta _{xxx}$in the equation. This term is essential in two
respects: first it introduces the dispersion in a lower order of magnitude
than in the traditional case [1], and second, it is responsible for
dispersion in the case of the cylindrical geometry [10]. As it should be,
the generalized KdV equation and its formal general traveling solution
approach the KdV ones, in the shallow layer limit of the model.

Based on these results we conjectured that there may exists a deep
connection between NPDE, infinite order linear ODE and finite-difference
equation, together with their symmetries. In short, we have shown that by
starting from a one-dimensional model for an ideal incompressible
irrotational fluid layer, a more general differential structure thanthe KdV
equation is obtained and it has interesting properties, not all explored,
that reduce as they approach the KdV equation in the shallow layer limit.

We also stress that the result of section 3, namely the realization of the
gKdV equation in terms of a DFDE could be the starting point for searching
for more interesting symmetries. Finite-difference equations are related to
self-similarity and wavelet bases, too. The $B$ parameter in our model is
similar with the scale parameter in multiresolution analysis, and we can
obtain in this way a connection with this field.

We anticipate many possible applications of such a formalism, especially in
the field of clusters, fusion, fission, drops, bubbles and shells. If the
gKdV equation could be shown to arise from a Lagrangean, or Hamiltonian
formalism, one can apply this result in physical models which involve
nonlinear shapes or contours.

\bigskip 

REFERENCES

\bigskip

[1] \ A. Ludu and J. P. Draayer, {\it Phys. Rev. Lett}. {\bf 80} (1998)
2125; R. G. Holt and E. H. Trinh, {\it Phys. Rev. Lett.} {\bf 77} (1996)
1274; R. E. Apfel {\it et al}., {\it Phys. Rev. Lett}. {\bf 78} (1997) 1912.

[2] \ A. Bohr, {\it Fys. Medd. K. Dan. Vidensk. Selsk}.,{\bf \ 26} (1952)
14; R. Gherghescu, A. Ludu and J. P. Draayer, {\it J. Phys. G: Nucl. Part.}%
{\bf \ 27 }(2001) 63.

[3] \ N. J. Zabusky {\it et al,} {\it J. Comput. Phys}. {\bf 30} (1979) 96.

[4] \ E. A. Overman and N. J. Zabusky, {\it Phys. Rev. Lett.} {\bf 45}
(1980) 1693.

[5] A. T. Dorsey and R. E. Goldstein, {\it Phys. Rev. B} {\bf 57} (1998) 3059

[6] \ C. Wexler and A. T. Dorsey, {\it Phys. Rev. Lett.},{\bf \ 82} (1999)
620, {\it Phys. Rev. B}{\bf \ 60} (1999) 10971.

[7] \ L. Friedland and A. G. Shagalov, {\it Phys. Rev. Lett.}, {\bf 85}
(2000) 2941.

[8] M. Remoissenet, {\it Waves Called Solitons} (Springer-Verlag, New York,
1999); Y. Tsukuhara,{\it \ et al}, {\it Appl. Phys. Lett.}, {\bf 77} (2000)
2926.

[9] \ D. J. Kortweg and G. de Vries, {\it Phil. Mag.}, {\bf 39 }(1895) 422.

[10] \ R. Abraham and J. E. Marsden, {\it Foundations of Mechanics} (The
Bejamin/Cummings Publishing Company, Inc., Reading, Massachusets, 1978); G.
L. Lamb, {\it Elements of Soliton Theory} (John Wiley \& Sons, New York,
1980).

[11] \ G. B. Witham, {\it Linear and Nonlinear Waves} (John Wiley \& Sons,
New York, 1974).

[12] \ A. C. Scott, F. Y. F. Chiu and W. Mclaughlin, {\it Proc. IEEE}{\bf \
61} (1973) 1443.

[13] \ S. V. Vladimirov, M. Y. Yu and V. N. Tsytovich, {\it Phys. Rep.} {\bf %
241} (1994) 1.

[14] \ L. Bieberbach, {\it S. -B. Preuss. Acad. Wiss}. (1916) 940; K.
Loewner, {\it Math.Ann.},{\bf \ 89 }(1923) 103.

[15] \ L. de Branges, {\it Acta Math}., {\bf 154} (1985) 137.

\end{document}